\documentclass[twocolumn,showpacs,prb]{revtex4} 

\usepackage{graphicx}
\usepackage{dcolumn}
\usepackage{bm}

\begin{document}

\title{Quasi-particle spin susceptibility in heavy-fermion superconductors : An NMR study compared with specific heat results} 


\author{Hideki Tou}
 \email{tou@hiroshima-u.ac.jp}
\affiliation{%
Department of Quantum Matter, AdSM, Hiroshima University, Higashi-Hiroshima, 739-8530, Japan
}

\author{Kenji Ishida}
\email{kishida@scphys.kyoto-u.ac.jp}
\affiliation{%
Department of Physics, Graduate School of Science, Kyoto University, Kyoto, 606-8502, Japan}

\author{Yoshio Kitaoka}
\email{kitaoka@mp.es.osaka-u.ac.jp}
\affiliation{%
Department of Materials Science and Technology, Graduate School of Engineering Science, Osaka University, Toyonaka, Osaka 560-8531, Japan}

\date{\today}

\begin{abstract}
Quasi-particle spin susceptibility ($\chi^{qp}$) for various heavy-fermion (HF) superconductors are discussed on the basis of  the experimental results of NMR Knight shift ($K$), NMR relaxation rate ($1/T_1$), and  electronic specific heat ($\gamma_{el}$) within the framework of the Fermi liquid model for a Kramers doublet crystal electric field (CEF) ground state. $\chi_{\gamma}$ is calculated from the enhanced Sommerfeld coefficient $\gamma_{el}$, and $\chi_{T_1}$ from the quasi-particle Korringa relation $T_1T(K_{T_1})^2=const.$ via the relation of $\chi_{T_1}=(N_A\mu_B/A_{hf})K_{T_1}$ where $A_{hf}$ is the hyperfine coupling constant, $N_A$ the Abogadoro's number and $\mu_B$ the Bohr magneton. For the even-parity (spin-singlet)  superconductors CeCu$_2$Si$_2$, CeCoIn$_5$ and UPd$_2$Al$_3$, the fractional decrease in the Knight shift, $\delta K^{obs}=K^{obs}(T_c)-K^{obs}(T\rightarrow 0)$, below the superconducting transition temperature ($T_c$) is  due to the decrease of  the spin susceptibility of heavy quasi-particle estimated consistently from $\chi_{\gamma}$ and $\chi_{T_1}$. This result allows us to conclude that the heavy quasi-particles form the spin-singlet Cooper pairs in  CeCu$_2$Si$_2$, CeCoIn$_5$ and UPd$_2$Al$_3$. On the other hand, no reduction in the Knight shift is observed in UPt$_3$ and UNi$_2$Al$_3$, nevertheless the estimated values of $\chi_{\gamma}$ and $\chi_{T_1}$ are large enough to be probed experimentally. The odd-parity superconductivity is therefore concluded in these compounds. The NMR Knight shift result provides  a convincing way to classify the HF superconductors into either even- or odd-parity pairing. 
\end{abstract}
\vspace*{3mm}

\maketitle 

\section{Introduction}
Extensive experimental and theoretical works for heavy-fermion (HF) systems have uncovered  characteristic feature that a localized $f$ state at high temperatures crosses over to a delocalized HF one at low temperatures through the hybridization with conduction electrons. This occurs below a so-called {\it coherent Kondo temperature} $T_K$, being  compatible with a renormalized heavy quasi-particle band width, leading to rich emergent  phenomena such as either even- or odd-parity unconventional HF superconductivity (SC), anomalous magnetic or multiple ordering, etc.  It has been confirmed that most physical quantities are described  in terms of the Fermi-liquid theory, \cite{Yamada} e.g. revealing the enhanced Pauli magnetic susceptibility ($\chi(T)\simeq \chi(0)$), the temperature ($T$)-linear coefficient in specific heat  ($C=\gamma_{el} T$) where $\gamma_{el}$ is the enhanced Sommerfeld coefficient, the $T$-square behavior in resistivity ($\rho=\rho_0+AT^2$), etc.  

As listed in Table I, the HF SC has been found in cerium (Ce) and uranium (U) based HF compounds, such as CeCu$_2$Si$_2$,\cite{Steglich} CeIrIn$_5$,\cite{Petrovic1}, CeCoIn$_5$\cite{Petrovic2}, UBe$_{13}$, \cite{Ott,Ott2} UPt$_3$,\cite{Stewart,Franse,Kimura} URu$_2$Si$_2$,\cite{Palstra,Schlablitz,Maple} UNi$_2$Al$_3$,\cite{Geibel1,Sato} UPd$_2$Al$_3$,\cite{Geibel2,Grauel,Visser} at ambient pressure. In these heavy-fermion superconductors (HFS's), both large $\gamma_{el}$ value and specific heat jump, $\Delta C/C_{el}(T_{c})\simeq 1$, associated with the superconducting transition give  unambiguous evidence that the heavy quasi-particle itself takes part in the formation of the Cooper pairs. 

In recent years, an intimate interplay between antiferromagnetism (AFM) and SC has been the most interesting and outstanding issue in Ce-based HF systems.  It is believed that the SC in CeCu$_2$Si$_2$,\cite{Steglich} CeIrIn$_5$,\cite{Petrovic1} and CeCoIn$_5$\cite{Petrovic2} emerges at the border of AFM even at ambient pressure ($P = 0$). The finding of pressure-induced SC in CeCu$_2$Ge$_2$,\cite{Jaccard92} CePd$_2$Si$_2$,\cite{Grosche}  CeIn$_3$,\cite{Mathur} and CeRhIn$_5$\cite{Hegger} strongly suggest that AFM and SC are related to each other because $P$-induced SC occurs either when AFM vanishes or coexists with it. \cite{Mito}

It is well known that Knight-shift measurements played vital role for establishing  the Bardeen-Cooper-Schriefer (BCS) theory for spin-singlet s-wave superconductors,\cite{Yosida} and are the most powerful tool to identify which  odd- or even-parity Cooper pairing state is realized  in superconductors.  When the system undergoes a superconducting transition, the {\it spin} susceptibility $\chi_{s}$ does decrease to zero at $T$=0 below $T_c$ for even-parity (spin-singlet) superconductors as follows:
\begin{equation}
\chi_{s} = 2\mu_{B}^{2}N_{0}Y(T), 
\label{eq:chiBCS}
\end{equation}
where $Y(T)$ is the {\it Yosida function} defined by \cite{Leggett}
\begin{equation}
Y(T) = -\frac{2}{N_{0}}\int_{0}^{\infty} N_{BCS}(\varepsilon) \frac{df(\varepsilon)}{d\varepsilon}d\varepsilon,
\label{eq:Yosida}
\end{equation}
and $N_0$ is the density of states (DOS) at the Fermi level in the normal state, $N_{BCS}(\varepsilon)$ is the DOS in the BCS superconducting state, and $f(\varepsilon)$ is the Fermi-Dirac function.

In transition metals and alloys, the electronic state  is formed as follows. The crystal electric field (CEF) splitting is much larger than the energy scale of the intra-atomic spin-orbit coupling (SOC), the {\it orbital} part of susceptibility arises from the second order interband mixing effect with other bands,  giving rise to $T$ independent Van Vleck susceptibility. Accordingly, only its {\it spin} part yields the $T$ dependence of susceptibility. Since both the Knight shift ($K$) and the susceptibility ($\chi$) depend on $T$, the contributions from the {\it spin} and the {\it orbital} parts are decomposed by taking the Clogston-Jaccarino plot of $K(T)$ vs. $\chi(T)$ with $T$ as an implicit parameter. \cite{Clogston} In the superconducting state, the spin part of the susceptibility decreases to zero  in accordance with eqs.~(\ref{eq:chiBCS}) and (\ref{eq:Yosida}).

On the other hand,  the $f$-electron systems with the strong SOC are generally described  by a total angular momentum, $J=L+S$. Furthermore, the CEF splits the lowest $J$ manifold into several doublets and/or singlets. If the simple localized $f$ electron picture can be applied for HF systems, the classical Van Vleck susceptibility between the lowest lying CEF level forming quasi-particle bands and other CEF levels are given by
\begin{equation}
\chi_{VV}=2N_A\frac{g_J^2\mu_B^2}{Z}\sum_{\alpha,\beta}\frac{|\langle \alpha|J_i|\beta \rangle|^2}{E_{\beta}-E_{\alpha}},
\label{eq:VV}
\end{equation}
where $N_A$ the Abogadoro's number, $g_J$ the Lande factor, $\mu_B$ the Bohr magneton, $Z$ the distribution function.  $E_{\alpha}$ and $E_{\beta}$ are ground ($\alpha$) and excited ($\beta$) state energies. Note that this interband contribution reveals a {$T$ dependence} when the CEF splitting is compatible with $T$. That is, the $T$ dependence of both $K$ and $\chi$ are not due to  {\it real spin}, but to {\it fictitious spin} including the interband contributions. Therefore, it is difficult to conclude  the odd-parity superconducting state in case of the Van Vleck contribution being dominant,\cite{Norman0,Norman,Park} even though $K$ unchanges across $T_c$.  Anyway, in the HF systems, it is not so clear to extract the quasi-particle {\it spin} susceptibility $\chi^{qp}$ via the Clogston-Jaccarino plot, and it is crucially important to reconfirm the  $\chi^{qp}$ due to heavy quasi-particles.

\begin{table*}[htbp]
\caption{Physical quantities and NMR parameters for various HFS's. $T_K$, $\gamma_{el}$, $T_c$, $A_{hf}^{i}$, and $(1/T_1T)_{T_c,i}$ are Coherent Kondo temperature, Sommerfeld coefficient (at $T_c$), superconducting transition temperature, hyperfine coupling constant, NMR spin-lattice relaxation rate divided by temperature at $T_c$, respectively. Here $i=\parallel, \perp$ (see text).}
{\small
\begin{center}
\begin{tabular}{l|cccccccc}
            & $T_K$  & $\gamma_{el}$ & $T_c$ & orientation & $\chi_i(T_c)$  &$A_{hf}^i$ & $(1/T_1T)_{T_c,i}$ & nucleus  \\ 
            & (K)      & (mJ/moleK$^2$)  & (K) & & (10$^{-2}$emu/mol) &(kOe/$\mu_B$) & (1/secK) &  \\ \hline 
CeCu$_2$Si$_2$ [1]& $\approx$10 & 1000 & 0.6      & $\parallel$ & 2 & $-$4.2    & 13.3  & $^{63}$Cu \\ 
               &          &      &          & $\perp$     & 1.6 & 1.5     & 77    & $^{63}$Cu \\ 
               &          &      &          & $\parallel$ & & $-$1.9    & 11.4  & $^{29}$Si \\ 
               &          &      &          & $\perp$     & & 6.4     & 9.6   & $^{29}$Si \\ 
UBe$_{13}$  [2]   & $\approx$10 & 1100 & 1        & Average         & 1.5 & 0.48    & 0.2   & $^{9}$Be(II) \\
UPt$_3$ [3]       & $\approx$10 & 420  & 0.55/0.5 & $\parallel$  & 0.45 & $-$70.9   & 1575  & $^{195}$Pt \\
               &          &      &          & $\perp$      & 0.85 & $-$84.0   & 1050  & $^{195}$Pt \\
URu$_2$Si$_2$ [4] & $\approx60$ & 65.5 &   1.2    & $\parallel$  & 0.46 & 3.6     & 0.012 & $^{29}$Si \\
               &          &      &          & $\perp$      & 0.15 &3.6     & 0.047 & $^{29}$Si \\
UPd$_2$Al$_3$ [5] & $\approx80$ & 150  &   2      & $\parallel$  & 0.55 &3.5     & 0.32  & $^{27}$Al \\
               &          &      &          & $\perp$      & 1.25 &3.5     & 0.25  & $^{27}$Al \\
UNi$_2$Al$_3$ [6] & $>300$?  & 120  &   1      & $\parallel$  & 0.35 &4.2     & 0.4  & $^{27}$Al \\
               &          &      &          & $\perp$       &0.15 &4.2     & 0.8  & $^{27}$Al \\
CeCoIn$_5$  [7]   & $\approx20$?& 350  &    2.3   & $\parallel$  & 1.2 &---    & 105  & $^{115}$In \\
               &          &      &          & $\perp$       & 0.7 &10.3$\sim$12.08    & ---  & $^{115}$In \\ \hline
Sr$_2$RuO$_4$ [8] & ($T^*\approx200$)\footnotemark[9]& 39   &    1.5   & $\perp$  &  0.095     & $-$250    & 15   & $^{99}$Ru \\          
\end{tabular}
\end{center}
}
\medskip
[1]{from Refs.~\cite{Steglich,Horn,Kitaoka,Ohama} }
[2]{from Refs.~\cite{Ott,Ott2,Tien}}
[3]{from Refs.~\cite{Stewart,Franse,Kimura,Tou2,Tou3,Vithayathil,Lee}}
[4]{from Refs.~\cite{Palstra,Schlablitz,Maple,Kohori3}}
[5]{from Refs.~\cite{Geibel1,Visser,Kyogaku,Tou1,Kohori1}}
[6]{from Refs.~\cite{Geibel2,Sato,Ishida1}}
[7]{from Refs.~\cite{Petrovic2,Curro,Kohori2}}
[8]{from Refs.~\cite{Maeno,IshidaSr,IshidaRu}}
[9]{taken from the metallic behavior in resistivity (Refs.~\cite{Maeno})}
\end{table*}

In this paper, we show that  the quasi-particle spin susceptibility $\chi^{qp}$ in HFS's is reasonably  estimated from the NMR and specific heat results on the basis of the Fermi-liquid theory with a Kramers doublet CEF ground state. \cite{Yamada} In {\S} 2, we remark that $\chi^{qp}$ is independently estimated from the Sommerfeld coefficient $\gamma_{el}$ and the NMR relaxation rate $1/T_1$, which are consistent with $\chi^{qp}$ estimated from the Clogston-Jaccarino plot. In {\S} 3, we describe the basic assumptions for the present analysis. Section 4 is devoted to the analysis of the experimental data for various HFS's in terms of the Fermi-liquid model. We also discuss the validity of the present model. After the brief comments on the quasi-particle susceptibility in the superconducting state in {\S} 5, we discuss the relation between the Knight shift and the parity of the order parameter (OP) in the superconducting state in {\S} 6.  From these analyses, CeCu$_2$Si$_2$, UPd$_2$Al$_3$, and CeCoIn$_5$ are reinforced to be an even-parity (spin-singlet) superconductor, whereas UPt$_3$ and UNi$_2$Al$_3$ are reconfirmed to be an odd-parity (spin-triplet) superconductor.  Finally, {\S } 7 is devoted to several concluding remarks.   

\section{NMR Parameters and Their Relation to Fermi-Liquid Parameters}
\subsection{Quasi-particle susceptibility}

In the Fermi-liquid theory for HF system, the electronic specific-heat coefficient $\gamma_{el}$ (in units per mole) and  the quasi-particle spin susceptibility $\chi^{qp}$ (in units per mole) are enhanced by  the mass enhancement factor $\tilde{\gamma}$, and   the magnetic enhancement factor $\tilde{\chi}_0$ associated with the contribution of $f$-electrons to the spin susceptibility, respectively. These are approximately written as  \cite{Yamada}
\begin{eqnarray}
 \gamma_{el}& \simeq & \frac{2}{3}N_A\pi^2k_B^2\rho^{f}(\varepsilon_F)\tilde{\gamma},
 \label{eq:gamma2}\\
\chi^{qp} & \simeq &\frac{2}{3}N_Ag_J^2J_{eff}^2\mu_B^2\rho^{f}(\varepsilon_F)\tilde{\chi}_0,
 \label{eq:chi2}
\end{eqnarray}
where $k_B$ is the Boltzmann factor, $J_{eff}$ the effective spin, $\rho^{f}(\varepsilon_F)$ the one-spin DOS at the Fermi level for bare $f$-electrons. 
Here, the DOS per spin of quasi-particles
 is approximately written by 
 \begin{equation}
 \rho^{*}(\varepsilon_F)\simeq\rho^{f}(\varepsilon_F)\tilde{\gamma}. 
 \label{eq:QPDOS}
\end{equation}
 From eqs.~(\ref{eq:gamma2})$-$(\ref{eq:QPDOS}), $\chi^{qp}$ is related to $\gamma_{el}$ as
\begin{eqnarray}
\chi^{qp}&=&\frac{2}{3}N_Ag_J^2J_{eff}^2\mu_B^2\rho^{\ast}(\varepsilon_F)\frac{\tilde{\chi}_0}{\tilde{\gamma}} \nonumber \\
 &=& \frac{\gamma_{el} g_J^2J_{eff}^2\mu_B^2}{\pi^2k_B^2}R,
 \label{eq:chi3}
\end{eqnarray}
where $R$ is the so-called {\it Wilson Ratio}, defined by $R\equiv\tilde{\chi}_0/\tilde{\gamma}$. 

When an external magnetic field is applied, $\chi^{qp}$ produces a small additional magnetic field at the nucleus. This gives rise to a Knight shift ($K^{qp}$) due to the quasi-particles that is scaled to $\chi^{qp}$ through the hyperfine-coupling constant $A_{hf}$ (in units per $\mu_B$) at ${\bf q}=0$  as
\begin{equation}
K^{qp} = \frac{A_{hf}}{N_A\mu_B}\chi^{qp}.
 \label{eq:Kchi}
\end{equation}
From eqs.~(\ref{eq:chi3}) and (\ref{eq:Kchi}), we obtain the quasi-particle spin contribution to the Knight shift, $K^{qp}$,  as  
\begin{equation}
K^{qp} = \frac{A_{hf}}{N_A\mu_B}\frac{\gamma_{el} g_J^2J_{eff}^2\mu_B^2}{\pi^2k_B^2}R.
 \label{eq:Kqpgamma}
\end{equation}
 Here it is noteworthy that {\it no interband contributions are included in this formula}. 

\subsection{Quasi-particle Korringa relation for isotropic system}
The nuclear spin-lattice relaxation process basically occurs through the spin-flip process between the state $|-1/2\rangle$ and $|+1/2\rangle$ of conduction electrons. Provided that HF state is realized at low temperatures below $T_K$,  $K^{qp}$ is also extracted using the NMR relaxation rate, $1/T_1$.  

In general, $1/T_1$ is given by \cite{Moriya}
\begin{equation}
\frac{1}{T_{1}}=\frac{2\gamma_{N}^{2}k_{B}T}{(g_J\mu_B)^{2}}\lim_{\omega\rightarrow0}\sum_{\bf q}\left|A_{hf}({\bf q}) \right|^2\frac{Im\chi_{\perp}({\bf q},\omega)}{\omega},
 \label{eq:generalT1}
\end{equation}
where $\gamma_N$ is the nuclear gyromagnetic ratio, $A_{hf}({\bf q})$ the ${\bf q}$-dependent hyperfine-coupling constant and $Im\chi_{\perp}({\bf q},\omega)$ the imaginary part of the transverse component of dynamical susceptibility $\chi({\bf q},\omega)$ of quasi-particles. If the HF state is formed above $T_c$,  the imaginary part of $\chi({\bf q},\omega)$ is expressed within the random phase approximation (RPA) as, \cite{Yamada,Kuramoto0}
\begin{equation} 
Im\chi({\bf q},\omega)\simeq \tilde{\chi}({\bf q},\omega)^2 Im\chi_0^f({\bf q},\omega),
 \label{eq:RPA}
\end{equation}
where $\chi_0^f({\bf q},\omega)$ is the dynamical susceptibility of $f$-electrons and $\tilde{\chi}({\bf q},\omega) $  an enhancement factor due to the electron-electron interaction as 
\begin{equation}
\tilde{\chi}({\bf q},\omega) \equiv  \left|\frac{\chi({\bf q}, \omega)}{\chi_0^f({\bf q}, \omega)}\right|.
 \label{eq:RPA2}
\end{equation}

 Since the imaginary part of $\chi_{0}^f({\bf q},\omega)$ near at the Fermi level $\varepsilon_F$ is expressed as \cite{Kuramoto0,Kuramoto}
%
%
\begin{equation}
Im \chi_{0}^f({\bf q},\omega) = \frac{2}{3}\pi g_J^2\mu_B^2 J_{eff}^2 \omega\sum_{\bf k}\left(-\frac{\partial f(\varepsilon)}{\partial \varepsilon _{\bf k}}\right)\delta (\omega -\varepsilon_{{\bf k}+{\bf q}}+\varepsilon_{{\bf k}}),  \label{eq:Imchi}
\end{equation}
%
%
then we have 
\begin{equation}
\lim_{\omega\rightarrow0}\sum_{\bf q}\frac{Im \chi({\bf q},\omega)}{\omega}=\frac{2}{3}\pi g_J^2\mu_B^2 J_{eff}^2\rho^{f}(\varepsilon_F)^{2}\left\langle\sum_{\bf q}\tilde{\chi}_{\bf q}^2\right\rangle,
 \label{eq:Imchi2}
\end{equation}
where $\tilde{\chi}_{\bf q} \equiv\tilde{\chi}({\bf q},\omega\rightarrow0)$ at the NMR frequency and the bracket represents averaging over the Fermi surface. If we are neglect the ${\bf q}$ dependence of the hyperfine coupling, then $A_{hf}({\bf q})$ is replaced by $A_{hf}$ and the NMR relaxation rate $1/T_1$ is expressed as
\begin{equation}
\frac{1}{T_{1}}=\frac{4}{3}\frac{\pi}{\hbar}(\gamma_N\hbar A_{hf})^{2}(g_JJ_{eff})^2\rho^{f}(\varepsilon_F)^{2}k_{B}T \left\langle\sum_{\bf q}\tilde{\chi}_{\bf q}^2\right\rangle,
 \label{eq:T1HF}
\end{equation}
From eqs.~(\ref{eq:chi2}), (\ref{eq:Kchi}) and (\ref{eq:T1HF}), we have
\begin{equation}
\frac{1}{T_{1}T}=\frac{3k_B\pi}{\hbar}\left(\frac{\gamma_N\hbar}{g_JJ_{eff}\mu_B}\right)^2(K^{qp})^2 {\cal K}(\alpha)
\label{eq:QPKorringa}
\end{equation}
where ${\cal K}(\alpha)\equiv \left\langle\sum_{\bf q}\tilde{\chi}_{\bf q}^2\right\rangle/\tilde{\chi}_0^2$ is the enhancement factor and $\alpha$ is the parameter associated with the Storner factor. \cite{Narath}  If the enhancement factor $\left\langle\sum_{\bf q}\tilde\chi_{\bf q}^2\right\rangle$ is the same as that at ${\bf q}=0$, {\it i.e}, ${\cal K}(\alpha)=1$, this relation is similar to the so-called {\it Korringa law}.  Then, when ferromagnetic and antiferromagnetic spin fluctuations become dominant due to electron-electron correlation effect,  ${\cal K}(\alpha)<1$ and  ${\cal K}(\alpha)>1$ are expected, respectively. Here we should notice that {\it $1/T_1$ does not include any interband contributions associated with the excitations across the CEF splitting}. This is  because $1/T_1$ involves the imaginary part of the dynamical susceptibility in the low energy limit as $\omega\rightarrow0$, as seen from eqs.~(\ref{eq:generalT1}) and (\ref{eq:Imchi2}). Anyway, $K^{qp}$ derived from the {\it quasi-particle Korringa relation} does indeed give the quasi-particle susceptibility.

\section{Basic Assumptions for Analysis}
\subsection{CEF ground state}

In Table I, we notice that the NMR $1/T_1$ for HFS's is enhanced by more than ten times in comparison with that for non-HF materials, \cite{Kitaoka,Kohori2,Kyogaku,Tou1,Kohori1,Tien,MacLaughlin,Vithayathil,Lee,Ishida1,Kohori3} for instance $1/T_1T=33$ (1/sK) for $^{195}$Pt of pure Pt metal\cite{Metallic}, and  $1/T_1T=0.033$ (1/sK) for $^{27}$Al for LaPd$_2$Al$_3$,\cite{Fujiwara} etc. This means that the heavy quasi-particles are involved in the NMR relaxation process. Since the non-Kramers ground state is magnetically intact, it is difficult to explain the enhanced relaxation rates observed in HFS's by the present Fermi-liquid model.  As mentioned in the previous section, the spin-flip process of conduction electrons would not occur through the Van Vleck type inter-band excitations in the low-energy limit as $\omega\rightarrow0$. Thus, we assume that the enhanced relaxation rate is ascribed to the magnetically degenerated HF ground state.   Namely, we assume that several doublets and singlets lying in lower energy region than $k_BT_K$ are renormalized into a ground state having Kramers degeneracy via the $c-f$ mixing effect.

\subsection{Effective moment below $T_K$}
In the spherical system without any CEF, the effective spin is defined as $J_{eff}^2=J(J+1)$ and the effective moment is $\mu_{eff}=g_J\sqrt{J(J+1)}\mu_B$. For the free electrons, $J_{eff}=\sqrt{3/2}$ and $\mu_{eff}=1.73\mu_B$. In HFS's, the CEF level scheme, that is determined experimentally, should be taken into account. Furthermore, several doublets and singlets are expected to be renormalized by the $c-f$ mixing effect in the heavy-Fermi-liquid state. It is, however, underlying issue to estimate to what extent the CEF level splitting affects $\chi^{qp}$ in HFS's. 

For Ce-based HFS's having the clear CEF level splitting, the basis function of a CEF ground state is not spherical but anisotropic. In order to introduce the anisotropy, we assume that the effective moment at low $T$'s is attributed to only the CEF ground state level and we do not take the average of each direction. Then, the effective moment is tentatively replaced by $\mu_{eff}^{i}=g_JJ_{eff}^{i}\mu_B=g_J\langle 0|J_{i}|0\rangle\mu_B$ where $|0\rangle$ is the basis function of a CEF ground state. For CeCu$_2$Si$_2$, using the CEF scheme,\cite{Horn} the respective effective moments at low $T$ for $H\perp c$ and $H\parallel c$ are estimated as $\mu_{eff}^{\perp}=g_J\mu_B\langle 0|J_{\perp}|0\rangle\simeq0.90\mu_B$ and $\mu_{eff}^{\parallel}=g_J\mu_B\langle 0|J_{\parallel}|0\rangle\simeq1.07\mu_B$. For CeCoIn$_5$,\cite{Curro} the effective moment for $H\perp c$ is also calculated as $\mu_{eff}^{\perp}\simeq1.28\mu_B$. 

By contrast,  for U-based systems, a clear CEF level splitting has not been reported yet. This has been generally accepted as the itinerant nature of U 5f-electrons. So that, the basis function would be rather spherical. In our analysis, the effective moment for U-based systems is tentatively taken as $\mu_{eff}=1.73\mu_B$, i.e., free electron value. The magnetic anisotropy is, however, expected from the various magnetic measurements. Therefore the effective moment for the magnetically easy axis is taken to be $\mu_{eff}^{easy}=1.73\mu_B$, while that for the magnetically hard axis is to be $\mu_{eff}^{hard}=1.73\mu_B\times\sqrt{\chi^{obs}_{hard}/\chi^{obs}_{easy}}$ where the ratio, $\chi^{obs}_{hard}/\chi^{obs}_{easy}$, is estimated from the anisotropy in magnetic susceptibility at $T_c$ listed in Table I.

\subsection{Hyperfine coupling constant}
The ${\bf q}$-dependent hyperfine coupling constant $A_{hf}({\bf q})$ in eq.~(\ref{eq:generalT1}) is the spatial Fourier transform of the hyperfine coupling $A_{hf}({\bf r})$ between nucleus and Ce/U ions separated a distance ${\bf r}$. If the nucleus is coupled to only one Ce/U ion, $A_{hf}({\bf q})$ is independent of ${\bf q}$. In the HFS's, the calculation of $A_{hf}({\bf q})$ would require one to sum over contributions from the next nearest Ce/U neighbors. However, $A_{hf}({\bf r})$ is not so simple. Therefore, under the most naive assumption, we neglect the wave number dependence of the hyperfine coupling constant, i.e., $A_{hf}({\bf q})=A_{hf}$.

\subsection{Anisotropic quasi-particle Korringa relation}
In HF systems, the magnetic anisotropy is inevitable even in the HF state because the HF state at low $T$ is formed through the $c-f$ hybridization. It should be noted that a large magnetic anisotropy in  $\chi^{qp}_i (i=a, b, c$-axes)  arises from the anisotropy in the hyperfine-coupling constant ($A_{hf}^{i}$), the effective moment ($\mu_{eff}^{i}$) and the enhancement factor ($\tilde{\chi}_0^{i}$). Unfortunately, no anisotropic Fermi liquid theory have been provided yet. Furthermore, the CEF splitting and the $c-f$ mixing effect are not so clear in quantitative level. Therefore, from a phenomenological point of view, a uniaxial magnetic anisotropy is taken into consideration  as
\begin{equation}
\chi^{qp}_{i}  =  \frac{2}{3}N_A{\mu_{eff}^{i}}^2\rho^{f}(\varepsilon_F)\tilde{\chi}_0^{i}, 
\label{eq:chianiso}
\end{equation}
where $i=\parallel$, and $\perp$ which mean the respective parallel and perpendicular components to the field direction, e.g., $\chi_{\parallel}=\chi_c$ and $\chi_{\perp}=\chi_a=\chi_b$. Then the Knight shift in eq.~(\ref{eq:Kchi}) or (\ref{eq:Kqpgamma}) is replaced by anisotropic Knight shift as 
\begin{eqnarray}
K_{i}^{qp}&=& \frac{A_{hf}^{i}}{N_A\mu_B}\chi^{qp}_{i}  \nonumber \\
    &= &\frac{A_{hf}^{i}}{N_A\mu_B}\frac{\gamma_{el} (\mu^{i}_{eff})^2}{\pi^2k_B^2}R_i.
\label{eq:kchianiso}
\end{eqnarray}

The quasi-particle Korringa relation, eq.~(\ref{eq:QPKorringa}), is modified using anisotropic parameters, $A_{hf}^{i}$, $\mu_{eff}^{i}$, and ${\cal K}_i(\alpha)$, as 
\begin{widetext}
\begin{equation}
 \left(\frac{1}{T_1T}\right)_\parallel = \frac{3\pi k_B}{\hbar} \left(\frac{\gamma_N\hbar}{\mu_{eff}^{\perp}} \right)^2(K_{\perp}^{qp})^2{\cal K}_{\perp}(\alpha), 
 \label{eq:Korringaparallel}
\end{equation}
and
\begin{equation}
\left(\frac{1}{T_1T}\right)_{\perp} =  \frac{3\pi k_B}{2\hbar} \left\{\left(\frac{\gamma_N\hbar}{\mu_{eff}^{\perp}} \right)^2 (K_{\perp}^{qp})^2{\cal K}_{\perp}(\alpha)+ \left(\frac{\gamma_N\hbar}{\mu_{eff}^{\parallel}} \right)^2 (K_{\parallel}^{qp})^2{\cal K}_{\parallel}(\alpha) \right\},
\label{eq:Korringaperp}
\end{equation}
\end{widetext}

\section{Discussion : the Normal State}
\subsection{Quasi-particle Korringa relation}

With above assumptions, we calculated the effective values of the quasi-particle Knight shifts at $T_c$ from eqs.~(\ref{eq:kchianiso}) $-$ (\ref{eq:Korringaperp}) by using parameters $\gamma_{el}$, $A_{hf}^i$, and $(1/T_1T)_{T_c,i}$ listed in Table I. Here we defined the effective quasi-particle Knight shift associated with $\gamma_{el}$ including the Wilson ratio $R$ as $K_{\gamma,i} \equiv K^{qp}_{i}/R_i$\, while that associated with $T_1$ including the enhancement factor ${\cal K}_i(\alpha)$ as $K_{T_1,i}\equiv K^{qp}_{i}\sqrt{{\cal K}_i(\alpha)}$.  Thus calculated values $K_{\gamma,i}$ and $K_{T_1,i}$ for various HFS's are listed in  Table II.

\begin{table*}[htbp]
\begin{center}
\caption{Observed reduction of the Knight shift $\delta K^{obs}\equiv K^{obs}(T_c)-K^{obs}(T\rightarrow 0)$ at $T\rightarrow 0$ K for various heavy fermion superconductors. $K_{\gamma}\equiv K^{qp}/R_{T_c}$ and $K_{T1}\equiv K^{qp}\sqrt{{\cal K}(\alpha)}$ are the calculated effective Knight shifts in the normal state from the electronic specific heat ($\gamma_{el}$) and the NMR relaxation rate ($1/T_1$) assuming $A_{hf}(q)=A_{hf}$, respectively.  Here we assume $\mu_{eff}\approx1.73\mu_{B}$ for U-based HFS's (see {\S} 3).  For UPt$_3$, the respective $\delta K^{obs}$'s for C and B-phases are listed.
}
{\small
\begin{tabular}{l|cccccc}
            & nucleus & orientation & $\delta K^{obs}_i$ & $K_{\gamma,i}$ & $K_{T_1,i}$  \\ 
            & & & (\%)      & (\%)  & (\%)    \\ \hline 
CeCu$_2$Si$_2$ [1] & $^{63}$Cu & $\parallel$  & $<-$0.2 & $-$0.39  & $-$1.41   \\ 
               & $^{63}$Cu & $\perp$     & 0.12  & 0.10  & 0.36  \\ 
               &$^{29}$Si & $\parallel$         & ---  & $-$0.18 & $-$0.57 \\ 
               &$^{29}$Si & $\perp$         & 0.45  & 0.42 & 0.42 \\ 

UBe$_{13}$ [2]    &$^{9}$Be(II)& Average & 0  & 0.13 & 0.163   \\
UPt$_3$ [3]       &$^{195}$Pt& $\parallel$  & 0/$-$0.16  & $-$3.89   & $-$2.89  \\
               &$^{195}$Pt& $\perp$  &  0/$-$0.15   & $-$8.75   & $-$10.20   \\
URu$_2$Si$_2$ [4] &$^{29}$Si& $\parallel$  & 0 & 0.057    & 0.074 \\
               & $^{29}$Si& $\perp$   &  0    &  0.019   & 0.016  \\
UPd$_2$Al$_3$ [5] &$^{27}$Al& $\parallel$  & 0.07  & 0.064  & 0.059  \\
               & $^{27}$Al& $\perp$     &  0.11 & 0.128  & 0.11   \\
UNi$_2$Al$_3$ [6] & $^{27}$Al& $\parallel$  & ---  &   0.053  & 0.129  \\
               & $^{27}$Al & $\perp$     &  0    &  0.123  & 0.124 \\
CeCoIn$_5$  [7]   &$^{115}$In & $\parallel$ &  0.5   &  ---    &  ---   \\
               & $^{115}$In & $\perp$    &  0.6   &  0.49    & 1.74  \\ \hline
Sr$_2$RuO$_4$ [8] & $^{99}$Ru  & $\perp$       &   0    & $-$2.4     & $-$3.87 \\          
\end{tabular}
}
\end{center}
[1]{from Refs.~\cite{Kitaoka} }
[2]{from Refs.~\cite{Tien}}
[3]{from Refs.~\cite{Tou2,Tou3}}
[4]{from Refs.~\cite{Kohori3}}
[5]{from Refs.~\cite{Kyogaku,Tou1,Kohori1}}
[6]{from Refs.~\cite{Ishida1}}
[7]{from Refs.~\cite{Kohori2}}
[8]{from Refs.~\cite{IshidaRu}}
\end{table*}

As clearly seen in the eqs.~(\ref{eq:kchianiso}) $=$ (\ref{eq:Korringaperp}), both $|K_{\gamma,i}|$ and $|K_{T_1,i}|$ depend on the size of $\mu_{eff}$. In order to check the validity of the present analysis, $|K_{T_1,i}|$ $(\equiv |K^{qp}_i|\sqrt{{\cal K}_i(\alpha)} )$ is plotted against $|K_{\gamma,i}|$ $(\equiv |K^{qp}_i|/R_{T_c} )$ for various HFS's in Fig. 1.  Most of values are consistent with a relation of $|K_{T_1,i}/K_{\gamma,i}|=R_{T_c}\sqrt{{\cal K}_i(\alpha)}=1$ except CeCu$_2$Si$_2$ and CeCoIn$_5$.

\begin{figure}[htbp]
\includegraphics[width=0.8\linewidth]{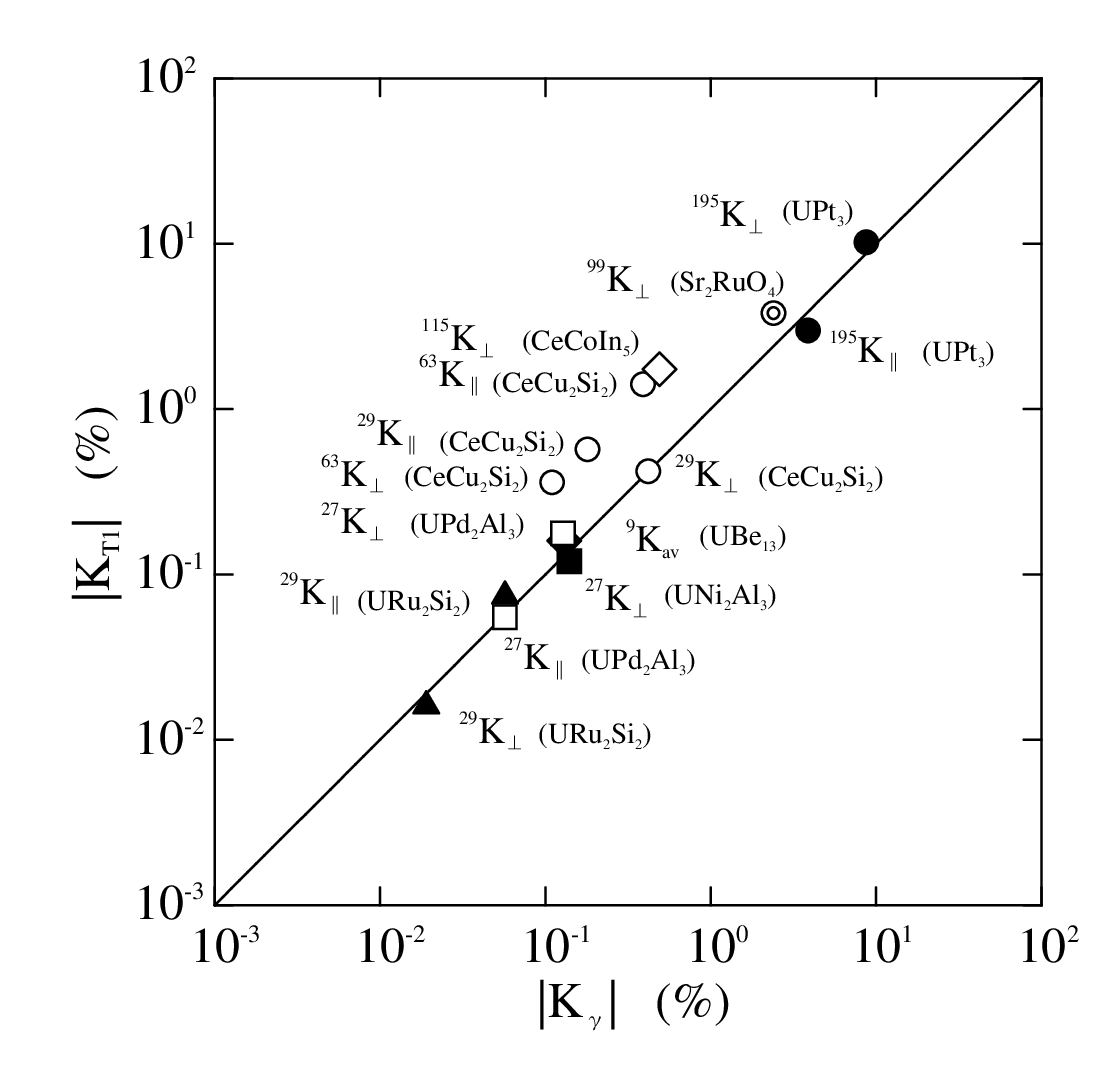}
\caption{ $|K_{T_1}|$ $(=|K^{qp}|\sqrt{{\cal K}(\alpha)} )$  plotted against $|K_{\gamma}|$ $(=|K^{qp}|/R )$ for various HFS's. $|K_{T_1}|$ and $|K_{\gamma}|$ are estimated from the quasi-particle Korringa relation ($T_1T(K^{qp})^2{\cal K}(\alpha)=constant$) and the electronic specific heat ($K^{qp}\propto\gamma_{el}R$) just above $T_c$, respectively. Here we assume $A_{hf}(q)=A_{hf}$ for calculations of $K_{T_1}$, and $\mu_{eff}\approx1.73\mu_{B}$ for U-based HFS's (for details,  see {\S} 3). The solid line shows $R\sqrt{{\cal K}(\alpha)}=1$.}
\end{figure}

According to the Fermi-liquid theory,\cite{Yamada,Kontani} the Wilson ratio varies in the range of 1…$R$…2. Simply, $R=1$ is the limit of electron-electron correlation parameter $U\rightarrow0$, while $R=2$ is the limit of $U\rightarrow\infty$. The recent theoretical studies have predicted that the mass enhancement factor is comparable to the magnetic one, {\it i.e.,} the Wilson ratio is $R\approx 1.2$.\cite{Cox,Kontani}  Actually in CeCu$_2$Si$_2$, under the present assumption of $\mu_{eff} (\approx 1\mu_B)$, we obtain $R_{T_c}=1.24$ from eq.~(\ref{eq:chi3}) with the experimental values, $\gamma_{el}\approx 1000$ mJ/(moleK$^2)$,\cite{Steglich} and $\chi^{obs}(T_c)\approx 6\times 10^{-3}$ (emu/mole).\cite{Lieke} A good agreement between the theoretical and the experimental values strongly suggests the Wilson ratio in the HF liquid state is on the order of unity.

For U-based HFS's, on the other hand, the absence of the clear CEF splitting prevents quantitative estimates for $\mu_{eff}$, consequently for $R_{T_c}$.  However $R_{T_c}$ would be more close to unity than that in Ce-based systems because U-based HFS's share more itinerant nature than Ce-based ones do. Thus the relation of $|K_{T_1,i}/K_{\gamma,i}|=R_{T_c}\sqrt{{\cal K}_i(\alpha)}=1$ suggests that $\mu_{eff}\approx1.73\mu_B$ is probably a good approximation for the effective moment in the U-based systems. Instead, if we assume $\mu_{eff}=1\mu_B$ and $R_{T_c}=1$ in U-based systems, we obtain ${\cal K}_i(\alpha)\approx 4$ that is comparable to that in Ce-based HFS's in which antiferromagnetic spin fluctuations are dominant and systems are close to a magnetic quantum critical point. \cite{Ishida2,Zheng} This is, however, unlikely to occur in U-based HFS's becasue $1/T_1T=const.$ behavior strongly suggests that antiferromagnetic spin fluctuations are not dominant at low $T$'s. 

To be emphasized here is that the quasi-particle Korringa relation is actually hold in the HFS's, though it is still unclear to evaluate $R_{T_c}$ and ${\cal K}_i(\alpha)$ experimentally. We also calculated the susceptibilities $\chi_{\gamma,i}(=\chi^{qp}_i/R_i)$ and $\chi_{T_1,i}(=\chi^{qp}_i\sqrt{{\cal K}_i(\alpha)})$ at $T_c$ by using eq.~(\ref{eq:kchianiso}) (see Table III).

\subsection{Van Vleck susceptibility}
Here we deal with the Van Vleck susceptibility $\chi_{VV,i}$. According to the theoretical arguments, \cite{Yamada,Zhang,Cox,Kontani} $\chi_{VV,i}$ is generally expressed as,
\begin{equation}
\chi_{VV,i}=\chi_{T,i}-\chi^{qp}_i,
\label{eq:chivvcal}
\end{equation}
where $\chi_{T,i}$ is the total susceptibility and  $\chi_{T,i}=\chi_i^{obs}(T_c)$ at $T_c$.  As discussed in the previous section, both $\chi_{\gamma,i}$ and $\chi_{T_1,i}$ provide the quasi-particle susceptibility $\chi^{qp}_i/R$ and $\chi^{qp}_i\sqrt{{\cal K}_i(\alpha)}$. If we assume $R_{T_c}\approx1$, $\chi^{qp}_i\approx\chi_{\gamma,i}$. That is, the Van Vleck susceptibility is approximately written as $ \chi_{VV,i}\simeq\chi_i^{obs}(T_c)-\chi_{\gamma,i}$.\cite{tou06} The calculated values of $\chi_{VV,i}$ are listed in Table III. Note that $\chi_i(T_c)$'s are larger than $\chi_{\gamma,i}$ and $\chi_{T_1,i}$ except for UBe$_{13}$,\cite{UBe13} and $\chi_{VV,i}$'s are comparable with or larger than the quasi-particle susceptibility. This seems to be consistent with the theoretical prediction that $\chi_{VV}$ can become comparable with or larger than the quasi-particle susceptibility ($\chi^{qp}$) via $c-f$ mixing. \cite{Norman0,Norman,Park,Kontani,Zhang,Cox} This in turn indicates the observed Knight shift $K^{obs}$ involves the Van Vleck contribution.

\section{Quasi-Particle Susceptibility in the Superconducting state}
Here, we discuss about a relation between the Knight shift and the parity of the superconducting order parameter (OP). 
In the superconducting state, the DOS  for quasi-particle excitations is written as \cite{Sigrist}
\begin{eqnarray}
\rho_{s}^{\ast}(\varepsilon) & = &  \int\frac{d\Omega_{\bf k}}{4\pi}\frac{\rho^{\ast}(\varepsilon_F)\left|\varepsilon\right|}{\sqrt{\varepsilon^{2}-\left|\Delta({\bf k})\right|^{2}}}, \nonumber \\
&=& \frac{\rho^{\ast}(\varepsilon_F)\left|\varepsilon\right|}{4\pi}\int_0^{2\pi}\int_0^{\pi}\frac{\sin\theta d\theta d\phi}{\sqrt{\varepsilon^{2}-\left|\Delta(\theta,\phi)\right|^{2}}},
\label{eq:SCDOS}
\end{eqnarray}
where the integral is over the solid angle of $\Omega_{\bf k}$ and $\rho^{\ast}(\varepsilon_F)$ is the quasi-particle DOS at the Fermi surface in the normal state defined by eq.~(\ref{eq:QPDOS}). Here $\Delta({\bf k})$ is the gap function and is given by
\begin{equation}
\Delta({\bf k})\propto \left\{
\begin{array}{ll}
\left|\Psi({\bf k})\right|  & (l=\mbox{even}), \\ 
\sqrt{\left|{\bf d}({\bf k})\right|^2\pm\left|{\bf d}({\bf k})\times{\bf d}^{\ast}({\bf k})\right|} & (l=\mbox{odd}),
\end{array}
\right.
\label{eq:eigenvalue}
\end{equation}
where $\Psi({\bf k})$ is a single even function for a spin-singlet pairing state ($S=0$, $l=$even), whereas ${\bf d}=(d_x,d_y,d_z$) is a {\bf d}-vector for a spin-triplet pairing state ($S=1$, $l=$odd). Due to the ${\bf k}$ dependence of the OP, the superconducting energy gap vanishes at points and/or along lines on the Fermi surface. 

The quasi-particle spin susceptibility, $\chi^{qp}$, below $T_c$ for even-parity superconductors is expressed as, 
\begin{equation}
\chi^{qp}_i = \frac{2}{3}N_A{\mu_{eff}^{i}}^{2}\rho^{\ast}(\varepsilon_F)R_i Y(T),
\label{eq:chisceven}
\end{equation}
and 
\begin{equation}
Y(T) = -\frac{2}{\rho^{\ast}(\varepsilon_F)}\int_{0}^{\infty} \rho^{\ast}_{s}(\varepsilon) \frac{df(\varepsilon)}{d\varepsilon}d\varepsilon.
\label{eq:Yosida2}
\end{equation}
Thus $\chi^{qp}$ decreases down to zero as $T\rightarrow 0$ for the even-parity (spin-singlet) superconductors, when the surface- and/or impurity effect are absent. Notice here that the anisotropy of the effective moment $\mu_{eff}^i$ causes the anisotropic decrease of $\chi^{qp}$ below $T_{c}$, depending on a field direction.

Differently from an even-parity pairing state, in an odd-parity pairing state, the Zeeman interaction, $(1/2)\chi^{qp}|{\bf H}\cdot{\bf d}|^2$, due to the spin of the Cooper pairs (${\bf d}$-vector) should be taken into consideration. Unless ${\bf d}$-vector is locked to the lattice, the application of field makes the {\bf d}-vector rotate as {\bf d}$\perp${\bf H} so as to minimize the Zeeman energy. In this case the quantization axis for the spin of the Cooper pairs is always parallel to the magnetic field. Namely, the quasi-particle susceptibility $\chi^{qp}$  is essentially the same as that in the normal state; $\chi^{qp}$ does not change below $T_c$. If the spin of the Cooper pairs is  ``{\bf frozen}'' with the lattice, \cite{Anderson1,Sigrist} it is expected that the quasi-particle susceptibility, $\chi^{qp}(T)$, decreases when ${\bf H}\parallel{\bf d}$ even for an odd-parity pairing state. It has been argued that the strong intra-atomic SOC may make ${\bf d}$ vector freeze. But, unfortunately, the inter-atomic SOC for the pairs remains quite obscure. \cite{Miyake}

\section{Discussion: the Superconducting State}
\subsection{Knight shift and parity of the Cooper pairs}
Figure 2 shows the $T$ dependence of the Knight shift for CeCu$_2$Si$_2$,\cite{Kitaoka} CeCoIn$_5$,\cite{Kohori2} and UPd$_2$Al$_3$.\cite{Kyogaku,Tou1} Figure 3 shows the $T$ dependence of the Knight shift for UPt$_3$ (for the C-phase in multiple phases) \cite{Tou2,Tou3}, UNi$_2$Al$_3$ \cite{Ishida1}, and Sr$_2$RuO$_4$.\cite{IshidaRu}  As clearly seen in Fig. 2, the reduction of the Knight shift below $T_c$ defined by 
\begin{equation}
\delta K^{obs}\equiv K^{obs}(T_c)-K^{obs}(T\rightarrow 0),
\label{eq:Kreduction}
\end{equation}
 is independent of the crystal direction for CeCu$_2$Si$_2$, CeCoIn$_5$, and UPd$_2$Al$_3$.\cite{Kitaoka,Kohori2,Kyogaku,Tou1} The fractional decrease in $K^{obs}$ below $T_c$ is ascribed to the reduction of $\chi^{qp}$ due to the formation of the spin-singlet Cooper pairs. In Fig. 3, on the other hand, the $\delta K^{obs}$'s  unchange across $T_c$ for UPt$_3$ and UNi$_2$Al$_3$, and also for Sr$_2$RuO$_4$.\cite{Tou2,Tou3,Ishida1}


\begin{figure}[htbp]
\centering{
\includegraphics[width=0.8\linewidth]{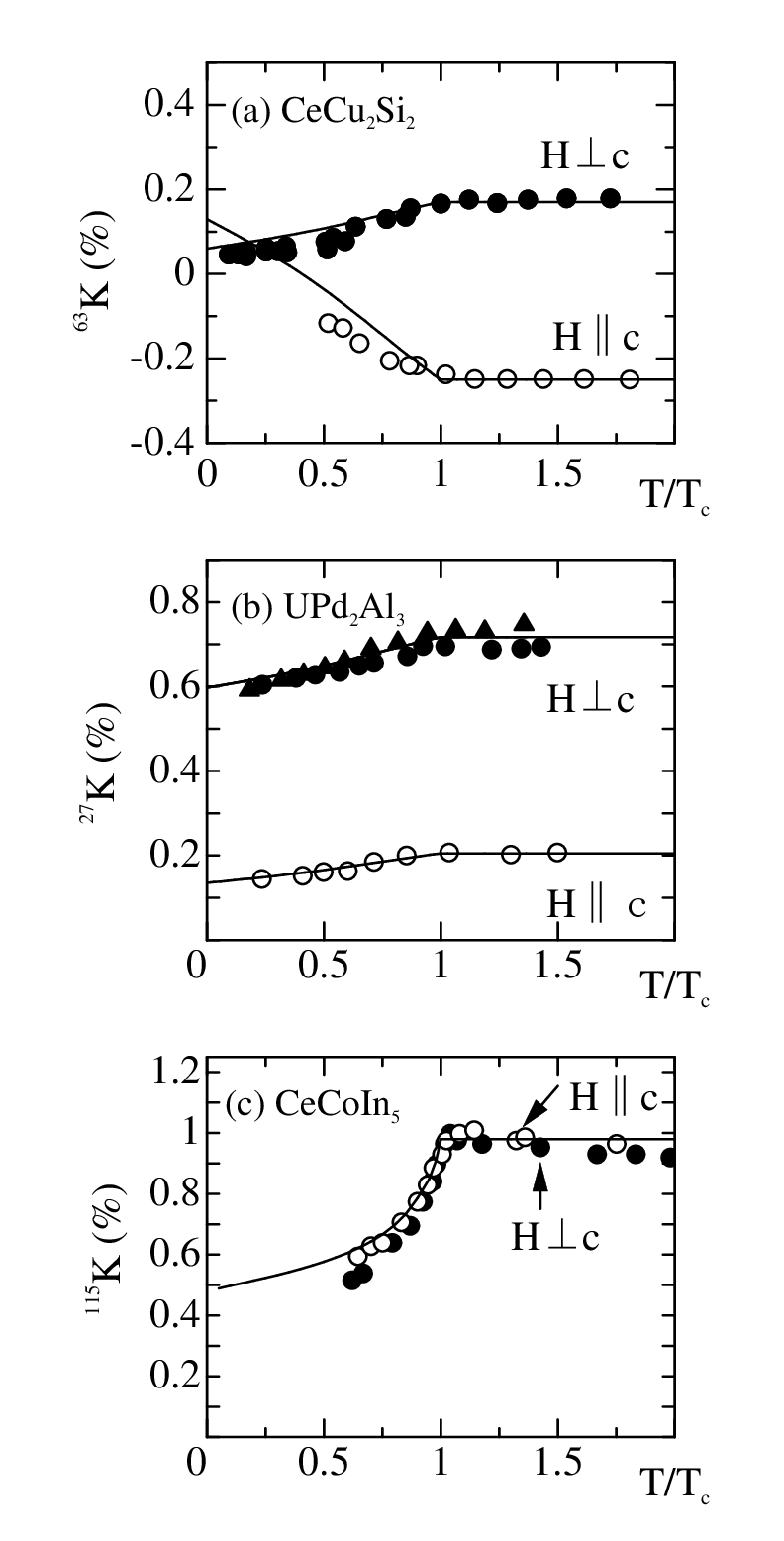}
}
\caption{
(a) $T$ dependence of the $^{63}$Cu-Knight shifts for oriented powder sample measured by Ueda {\it et al.}(Ref.~\cite{Kitaoka}). (b)  $T$ dependence of the $^{27}$Al-Knight shifts measured by Kyogaku {\it et al.} (Refs.~\cite{Kyogaku,Tou1,Kohori1}). Closed triangle shows the Knight shift for $H\perp c$ in oriented powder sample. Closed and open circles show the respective Knight shifts for $H\perp c$ and $\parallel c$ in single crystalline sample. (c) $T$ dependence of the $^{115}$In-Knight shifts for oriented powder sample measured by Kohori {\it et al.} (Ref.~\cite{Kohori2}). The solid lines in figures are calculations using the Sommerfeld coefficient $\gamma_{el}$ for a $d$-wave singlet $E_{1g}$ model (see {\S} 6.1). Here we assume $\mu_{eff}\approx1.73\mu_{B}$ for U-based HFS's (for details,  see {\S} 3).}
\end{figure}


The solid lines in the figures are calculated from eqs.~(\ref{eq:Kchi}), (\ref{eq:SCDOS}), (\ref{eq:chisceven}), and (\ref{eq:Yosida2}), with  the anisotropic energy gap $\Delta(T,{\bf k})$.  In our calculations, we tentatively choose the gap function as $d$-wave $E_{1g}$,\cite{Sigrist,Volovik}
\begin{eqnarray}
\Delta(T,{\bf k})&=&\Delta(T)(k_x\pm ik_y)k_z \nonumber \\
 &=& \Delta(T)\sin\theta\cos\theta \mbox{e}^{\pm i\phi}
\label{eq:delta}
\end{eqnarray}
with the $T$ dependence of the gap which is the same as the BCS theory as 
\begin{equation}
\Delta(T)=\Delta(0)\tanh\left[\frac{\pi k_BT_c}{\Delta(0)}\sqrt{a\frac{\Delta C}{C}\left(\frac{T_c}{T}-1 \right)} \right],
\label{eq:delta2}
\end{equation}
where $\Delta C/C$ is the specific heat jump, and $a=2$, and $\Delta(0)$ is the magnitude of the superconducting energy gap. Since the energy gap for the $E_{1g}$ symmetry vanishes along the line on the Fermi surface, {\it i.e.,} $\rho^{\ast}_s(\varepsilon)\propto \varepsilon$ ($\varepsilon\ll \Delta(0)$),  it can explain the $T^3$ behavior of NMR relaxation rate observed in the HFS's, where $1/T_1\propto \int_0^\infty\rho^{\ast}_s(\varepsilon)^2f(\varepsilon)(1-f(\varepsilon)) d\varepsilon \propto T^3$ for line-node gapped superconductors. Furthermore, we use $K_{\gamma}$ as $K^{qp}$ assuming $R_{T_c}=1$. This is because $K_{T_1}$ is sometimes enhanced by $\cal{K}(\alpha)$  for $R_{T_c}=1$, so that the spin part may be overestimated.

The calculations reproduce the observed fractional decrease in $K^{obs}$, i.e. $\delta K^{obs}_{\perp}\approx 0.12$ \% and $\delta K^{obs}_{\parallel}<-0.2$ \%  for CeCu$_2$Si$_2$ ($^{63}$Cu NMR) \cite{Kitaoka};  $\delta K^{obs}\approx 0.08\sim 0.12$ \% for UPd$_2$Al$_3$ ($^{27}$Al NMR)\cite{Kyogaku,Touex1}; $\delta K^{obs}\approx 0.5\sim 0.6$ \% for  CeCoIn$_5$ ($^{115}$In NMR).\cite{Kohori2} Here, the $T$ dependence of calculated Knight shift was obtained by using the appropriate energy gaps of $2\Delta(0)/k_BT_c=5$, $4$, and $7$ for CeCu$_2$Si$_2$, UPd$_2$Al$_3$, and CeCoIn$_5$, respectively.  Anyway, it is demonstrated that $\chi^{qp}$'s decrease to zero  as $T\rightarrow0$ due to the formation of  {\it spin-singlet} Cooper pairs in CeCu$_2$Si$_2$, CeCoIn$_5$, and UPd$_2$Al$_3$.  As a matter of fact, $\chi_{\gamma}$ (also  $\chi_{T_1}$) provides a reliable way to deduce $\chi^{qp}$ in the HF state. 

\begin{figure}[htbp]
\centering{
\includegraphics[width=0.8\linewidth]{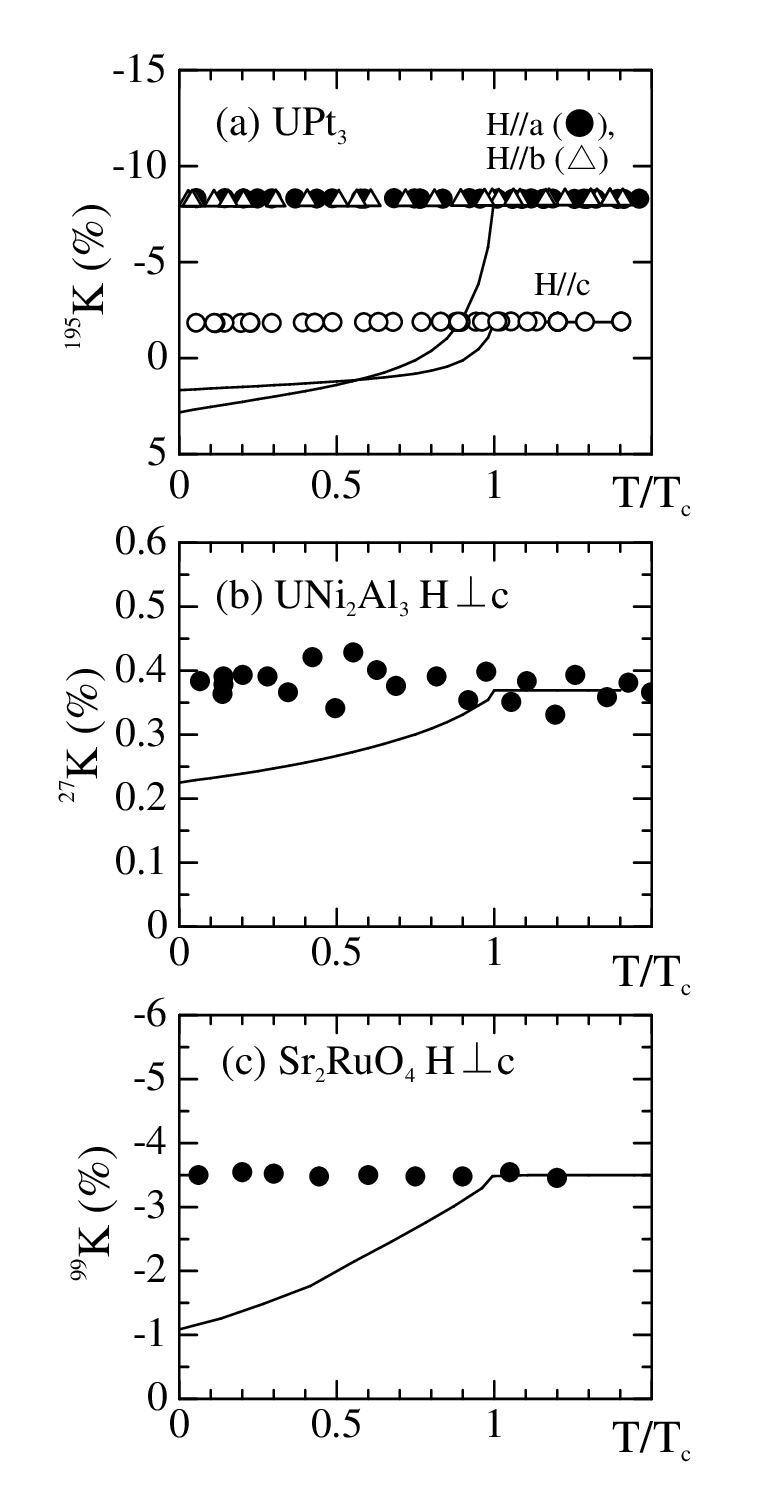}
}
\caption{(a) $T$ dependence of the $^{195}$Pt-Knight shifts for various field direction (Ref.~\cite{Tou2}). (b)  $T$ dependence of the $^{27}$Al-Knight shift for $H\perp c$-axis (Ref.~\cite{Ishida1}). (c)  $T$ dependence of the $^{99}$Ru-Knight shift for $H\perp c$-axis (Ref.~\cite{IshidaRu}). The solid lines in both figures are calculations using the Sommerfeld coefficient $\gamma_{el}$ for a $d$-wave singlet $E_{1g}$ model (see {\S} 6.1). Here we assume $\mu_{eff}\approx 1.73\mu_{B}$ for U-based HFS's (for details,  see {\S} 3).}
\end{figure}

Likewise, in Fig. 3, the $K_{\gamma}$ and $K_{T_1}$ in UPt$_3$ are estimated to be very large, which are compatible with the previous estimates both by the Clogston-Jaccarino plot and the high-$T$ Curie-Weiss fitting to the Knight shift. \cite{Tou2} Also those in UNi$_2$Al$_3$ are comparable to the values in UPd$_2$Al$_3$. Similar result is obtained for Sr$_2$RuO$_4$ in which a calculated value of $\chi_{\gamma}$ ($K_{\gamma}$) is in good agreement with the spin part of the susceptibility estimated from the Clogston-Jaccarino plot reported previously (See Tables II and III). \cite{IshidaRu} No significant reduction in $K^{obs}$ is observed in these compounds, ruling out a possibility for even-parity (spin-singlet) state. The solid lines in the figures are expected if a spin-singlet pairing states were realized with $2\Delta(0)/k_BT_c=$10, 5, and 4 for UPt$_3$,\cite{Kohori0} UNi$_2$Al$_3$,\cite{tou4} and Sr$_2$RuO$_4$,\cite{IshidaRu} respectively.  These experimental results reinforce that UPt$_3$, UNi$_2$Al$_3$, Sr$_2$RuO$_4$ are the odd-parity superconductors,  where the ${\bf d}$-vector rotates freely as ${\bf d}\perp{\bf H}$ in the C phase in UPt$_3$, \cite{Tou2} while it does as ${\bf d}\parallel c$-axis (the  odd-parity state with parallel spin pairing lying in the $ab$ plane) in UNi$_2$Al$_3$ \cite{Ishida1} and Sr$_2$RuO$_4$. \cite{IshidaRu}

\begin{figure}[htbp]
\centering{
\includegraphics[width=0.8\linewidth]{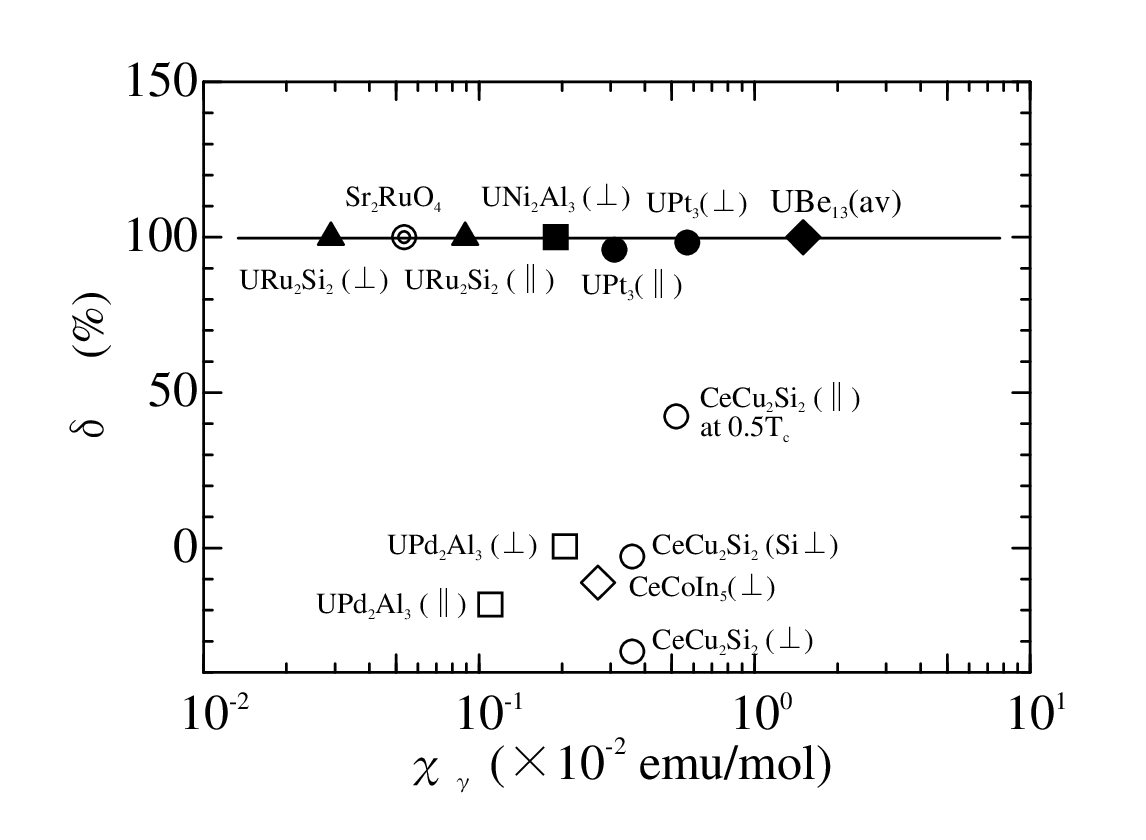}
}
\caption{ Normalized residual susceptibility at $T\rightarrow0$ defined as $\delta =(\chi_{\gamma}(T_c)-\delta\chi^{obs})/\chi_{\gamma}(T_c)$ plotted against $\chi_{\gamma}(T_c)$ which is related to the quasi-particle susceptibility as $\chi^{qp}(T_c)\simeq\chi_{\gamma}(T_c)$. For UPt$_3$, data points were calculated for the values in B-phase. 
}
\end{figure}

To make discussion more clear, we define the normalized residual susceptibility at $T\rightarrow 0$ as 
\begin{equation}
\delta \equiv \frac{\chi_{\gamma}(T_c)-\delta\chi^{obs}}{\chi_{\gamma}(T_c)}\simeq\frac{K_{\gamma}(T_c)-\delta K^{obs}}{K_{\gamma}(T_c)},
\end{equation}
where we assume $\delta K^{obs}=(A_{hf}/N_A\mu_B)\delta\chi^{obs}$. In Fig. 4, the residual susceptibility ratio $\delta$(\%)$=(\chi_{\gamma}(T_c)-\delta\chi^{obs})/\chi_{\gamma}(T_c)$ is plotted against $\chi_{\gamma}$. Note that  $\delta$ is almost zero (or below zero) for CeCu$_2$Si$_2$, CeCoIn$_5$, and UPd$_2$Al$_3$, ensuring that the $\chi^{qp}$ in these compounds decreases down to nearly zero at $T\rightarrow 0$ regardless of the crystal direction.  Together with the fact that $1/T_1\propto T^3$ below $T_c$, which excludes the $s$-wave BCS state,  a $d$-wave type anisotropic superconducting state is realized in CeCu$_2$Si$_2$, CeCoIn$_5$, and UPd$_2$Al$_3$.

By contrast, $\delta\approx 100$ (\%) for UPt$_3$ and UNi$_2$Al$_3$ gives evidence for an odd-parity (spin-triplet) superconducting state.  It was reported that  $K^{obs}$ unchanges across $T_c$ in UBe$_{13}$ \cite{Tien} and URu$_2$Si$_2$ as well. \cite{Kohori3} Although an odd-parity pairing state is most likely for URu$_2$Si$_2$ and UBe$_{13}$ from the present analyses, further precise measurements for high-quality single crystalline samples are desirable. Also the present analysis is applied to Sr$_2$RuO$_4$, reinforcing that it is the spin-triplet superconductor. \cite{IshidaSr,IshidaRu}

\subsection{Comments on quasi-particle susceptibility in UPt$_3$ }

Here we discuss the quasi-particle susceptibility in UPt$_3$. In the previous paper, \cite{Tou3} we reported that the value of $\delta \chi^{obs}_{b}\approx\delta\chi^{obs}_{c} \approx 1\times10^{-4}$ emu/mole is two orders of magnitude smaller than the value of measured susceptibility which is comparable with the present estimates, $\chi_{\gamma}$ and $\chi_{T_1}$. Fig. 5 shows the $T$ dependence of the $^{195}$Pt Knight shift ($^{195}K$) in UPt$_3$ at various magnetic fields. \cite{Tou3}  As seen in the figure, when the field strength is decreased,  the $K^{obs}_c$ for the $c$-axis (Fig. 5(c)) and  $K^{obs}_b$ for the $b$-axis (Fig. 5(b)) {\it decrease} across $T_c$ at $H<5$ kOe and $H<2.3$ kOe, respectively.  By contrast,  $K^{obs}_a$ (Fig. 5(a)) {\it unchanges} across $T_c$ down to the lowest field of $H=$ 1.764 kOe. Remarkably, $^{195}K$ depends on the strength of magnetic field and its crystal orientation. These results cannot be explained by the even-parity (spin-singlet) pairing model at all, but is consistent with the odd-parity (spin-triplet) one. \cite{Tou3}  

In order to discuss the quasi-particle susceptibility in UPt$_3$ within the present model, we focus again on the NMR relaxation behavior. It should be noted that  $1/T_1$ is enhanced in the HF state via the transverse component of dynamical magnetic susceptibility which is also responsible for the mass enhancement. \cite{Vithayathil,Lee} In most uranium compounds except UPd$_2$Al$_3$,\cite{Sato2} it is widely believed that a $5f^2$ configuration may lead to a non-Kramers ground state. For the HF state involving only a singlet CEF ground state, however, $1/T_1$ may not be enhanced, as mentioned in {\S} 3-1. The fact that $1/T_1$ is enhanced in HFS's suggests that the Kramers doublet or magnetic CEF ground state is responsible for the formation of HF state, associated with more itinerant nature of uranium 5$f$ electrons than of cerium 4$f$ electrons. In this context, the reason why the reduction of $\delta\chi^{obs}$ in the SC state is so small may be ascribed to the incomplete lock of ${\bf d}$-vector to the crystal, that is, the spin degree of freedom of the Cooper pairs is not always completely locked to the crystal axis. This suggests that the Knight shifts for $K_b^{obs}$ and $K_c^{obs}$ decrease markedly at further lower fields. The $^{195}$Knight shift measurements at fields lower than 1.7 kOe are needed to check this. \cite{kuber}

\begin{figure}[htbp]
\centering{
\includegraphics[width=1\linewidth]{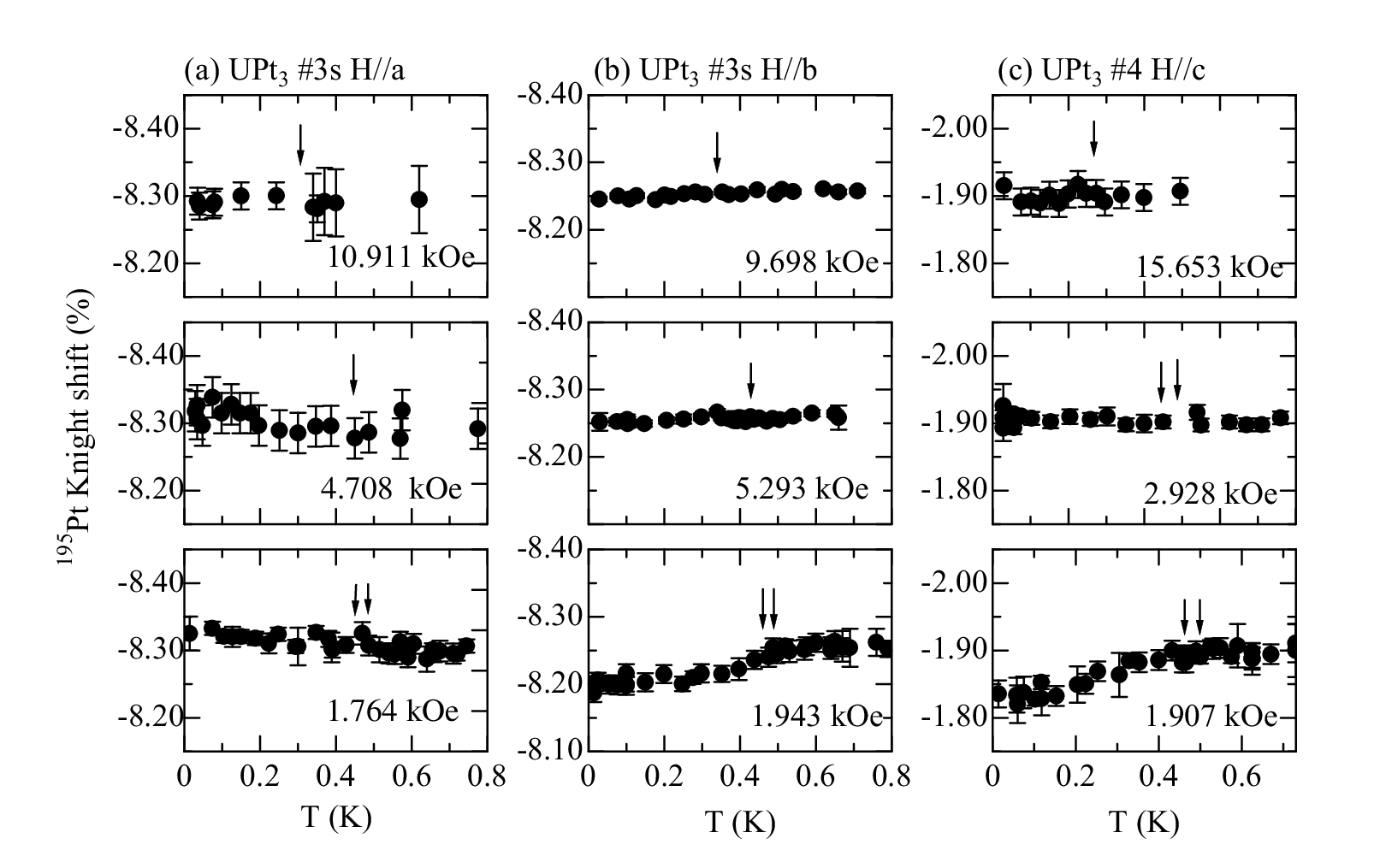}
}
\caption{ Temperature dependence of the $^{195}$Knight shift in UPt$_3$ at various magnetic fields for (a)$H\parallel a$, (b)$H\parallel b$, (c)$H\parallel c$ (Ref.~\cite{Tou3}). Arrows show $T_{c1}$ and $T_{c2}$. 
}
\end{figure}

Meanwhile, recent theoretical studies proposed that the origin of the tiny reduction of $\delta\chi^{qp}$ is possible to be explained by assuming that the U($5f^2$)-derived  non-Kramers singlet CEF ground state is hybridized with conduction electrons. \cite{Ikeda,Yotsuhasi} It was argued  that the $\chi^{qp}$ for a singlet CEF ground state is not enhanced, whereas the DOS of quasi-particles is enhanced via the on-site correlation effect giving rise to the heavy-Fermi liquid behavior. \cite{Ikeda}  It was predicted qualitatively that the NMR $1/T_1$ is enhanced  by an order of $1/T_K^2$, while $\chi^{qp}$ is not enhanced by correlation effect. \cite{Yotsuhasi}  Unfortunately no direct evidence for the CEF splitting in uranium compounds prevents the complete understanding of the HF state for the $5f^2$ configuration. Further quantitative discussions in theory in addition to experimental efforts are needed to test whether or not this theoretical model is applicable to UPt$_3$ as a realistic model. Anyway, the physical properties of the heavy-Fermion having a singlet CEF ground state should be addressed by both experimental and theoretical works in future. 

\begin{table*}[htbp]
\begin{center}
\caption{Reduction of the susceptibility $\delta \chi^{obs}=(N_A\mu_B/A_{hf})\delta K^{obs}$ at $T\rightarrow 0$K for various heavy fermion superconductors. The calculated effective susceptibility, $\chi_{\gamma}\equiv \chi^{qp}/R_{T_c}$ and $\chi_{T1}\equiv \chi^{qp}\sqrt{{\cal K}(\alpha)}$, by using the relation of $\chi_{\gamma,T_1}=(N_A\mu_B/A_{hf})K_{\gamma,T_1}$ assuming $A_{hf}(q)=A_{hf}$. Here we assume $\mu_{eff}\approx1.73\mu_{B}$ for U-based HFS's (see {\S} 3). For UPt$_3$, the respective $\delta\chi^{obs}$'s  for C and B-phases are listed. The Van Vleck susceptibility is calculated from $\chi_{VV}=\chi^{obs}(T_c)-\chi_{\gamma}$. $\chi_c$'s are the spin suceptibilities for pure metals. 
}
{\small
\begin{tabular}{l|ccccccccc}
            & nucleus & orientation &  $\delta \chi^{obs}_i$ & $\chi_{\gamma,i}$ & $\chi_{T_1,i}$ &$\chi_{VV,i}$ & $\chi_c$ & Parity \\ 
            & & &  (10$^{-2}$emu/mol)  &(10$^{-2}$emu/mol) & (10$^{-2}$emu/mol)& (10$^{-2}$emu/mol) & (10$^{-2}$emu/mol) & \\ \hline 
CeCu$_2$Si$_2$ & $^{63}$Cu& $\parallel$   & $>$0.26 & 0.52 & 1.88 & 1.43 & 0.001 & even \\ 
               & $^{63}$Cu & $\perp$      & 0.44 & 0.36 & 1.35 & 1.21 &  & \\ 
               & $^{29}$Si & $\parallel$  & --- & 0.52 &  1.6 & 1.43  & --- & \\ 
               & $^{29}$Si & $\perp$      & 0.37 & 0.36 &  0.36 & 1.21  &  & \\ 
UBe$_{13}$     & $^{9}$Be(II)& Average &  0    & 1.50 & 1.89 & 0.0  &  0.002 & odd(?) \\
UPt$_3$        & $^{195}$Pt& $\parallel$   &  0/0.0126 & 0.31   & 0.23 & 0.14  & 0.009 & odd \\
               & $^{195}$Pt& $\perp$   &0/0.0099 & 0.57  & 0.67 &0.28 & &  \\
URu$_2$Si$_2$  & $^{29}$Si& $\parallel$    &0  & 0.089 & 0.114 &0.371&--- & odd(?)\\
               & $^{29}$Si& $\perp$     & 0   & 0.029 & 0.025 &0.121& &  \\
UPd$_2$Al$_3$  & $^{27}$Al& $\parallel$   & 0.13    & 0.11  & 0.095 &0.15& 0.002 & even  \\
               & $^{27}$Al& $\perp$        & 0.204   & 0.205  & 0.177 &0.65& &  \\
UNi$_2$Al$_3$  & $^{27}$Al& $\parallel$    & ---  & 0.07  & 0.17 &0.118& 0.002 & odd  \\
               & $^{27}$Al& $\perp$        &  0   & 0.17  & 0.17 &0.34&  &  \\
CeCoIn$_5$     & $^{115}$In& $\parallel$     & --- & --- & ---&---& --- & even \\
               & $^{115}$In & $\perp$        & 0.32 & 0.27 & 0.94 &0.43&  & \\ \hline
Sr$_2$RuO$_4$  & $^{99}$Ru  & $\perp$            & 0   & 0.0535  & 0.068 &0.0415 &--- &odd \\    
\end{tabular}
}
\end{center}
\end{table*}

\section{Concluding remarks}
We have presented how the spin susceptibility $\chi^{qp}$ of quasi-particles in the HFS's is reasonably estimated within the framework of the Fermi-liquid theory, focusing on the NMR and specific-heat results. The present analysis for the Kramers doublet CEF ground state has provided two important results on the HF superconductivity: 
\begin{enumerate}
\item In the normal state, using the value of $T_1T$= const, and the $T$-linear coefficient $\gamma_{el}$ in specific heat, the respective values, $K_{T_1}$ and $K_\gamma$, are related to the quasi-particle Knight shift $K^{qp}$. The quasi-particle Korringa relation was found to be valid far below $T_K$ in the HFS's. For U-based HF superconductors, it  was expected that any significant wave-number dependence would be absent in low-lying excitations. This means that the on-sight correlation ($U$) might be responsible not only for the formation of the HF state but also for the occurrence of HF superconductivity. 

\item In the superconducting state, for CeCu$_2$Si$_2$, CeCoIn$_5$ and UPd$_2$Al$_3$, the spin part in the Knight shift deduced from $K_\gamma$ or $K_{T_1}$ is in excellent agreement with the reduction in the observed Knight shifts $\delta K^{obs}$ below $T_c$, giving unambiguous evidence that the heavy quasi-particles form the spin-singlet Cooper pairs. By contrast, based on the present analyses, UPt$_3$ and UNi$_2$Al$_3$ are concluded to belong to a class of odd-parity (spin-triplet)  pairing state. 

\end{enumerate}
The present NMR analysis has shed light on the semi-quantitative estimate for the quasi-particles susceptibility in the HF systems.  Especially from the present analysis based on the Fermi-liquid theory, it is suggested that the renormalized HF ground state has a Kramers degeneracy through the $c-f$ mixing even for U-based HFS's. However, we are still at a long way from a microscopic understanding how to represent the heavy-Fermi liquid state enhanced through the hybridization with a singlet CEF ground state. It is needed to be addressed by further experimental and theoretical efforts in future.

\begin{acknowledgements}
We are grateful to K. Asayama and G.-q. Zheng for useful discussions and for technical help with NMR measurements which were done at Osaka University. We are also grateful to N. Kimura, E. Yamamoto, Y. Haga, K. Maezawa, Y. \={O}nuki for providing the crystals of UPt$_3$, and to C. Geibel, F. Steglich, N. Aso, and N. K. Sato for providing CeCu$_2$Si$_2$, UPd$_2$Al$_3$ and UNi$_2$Al$_3$. Also we are grateful to Y. Maeno for providing Sr$_2$RuO$_4$. We would like to thank K. Miyake, H. Kohno, H. Ikeda, H. Kusunose, K. Yamada, and H. Kontani for theoretical comments and discussions. We thank J. Flouquet for valuable comments. H. T. thanks Y. Kohori for useful comments on URu$_2$Si$_2$ and CeCoIn$_5$, and M. Sera for helpful discussions. This work was supported by Grant-in-Aid for COE Reserach (Nos. 10CE2004 and 13CE2002) from the Ministry of Education, Sport, Science and Culture in Japan. 
\end{acknowledgements}

\newpage





\begin{references}
\bibitem{Konprivate} H. Kontani, and K. Yamada, private communications; The Van Vleck susceptibility discussed in Ref.\cite{Zhang,Cox,Kontani} includes the DOS of quasi-particles. However, this is an artificial factor, which cannot  be replaced simply by the quasi-particle DOS in the superconducting state.
\bibitem{UBe13} $^{9}$Be hyperfine field for Be(II) site is anisotropic, and anisotropic hyperfine fields were obtained as $A_{hf}^a=414$, $A_{hf}^b=147$, $A_{hf}^c=874$ Oe/$\mu_B$ for a single crystal UBe$_{13}$ [Ref.~\cite{clark}].  On the other hand, the $T$ dependence of NMR relaxation rate $1/T_1$ was measured for polycrystal UBe$_{13}$ [Ref.~\cite{MacLaughlin}]. In the present analysis, we tentatively used the average value of $A_{hf}$.
\bibitem{tou06} If we assume $R_{T_c}=2$, $\chi^{qp}$ becomes comparable to $\chi^{obs}(T_c)$, i.e., $\chi_{VV}\approx0$. This is hard to consider for the HFS's.
\end{references}
\end{document}